\documentstyle[12pt,psfig]{article}
\begin{document}

\title{Numerical and Theoretical Studies of Noise Effects in the
Kauffman Model}
\author{X. Qu\footnote{e-mail:quxh@uchicago.edu}, M. Aldana, Leo P. Kadanoff}
%\date{\today}
\maketitle
\begin{center}
The James Franck Institute, The University of Chicago.\\
5640 South Ellis Avenue, Chicago, Il, 60637.
\end{center}

\begin{abstract}
In this work we analyze the stochastic dynamics of the Kauffman model
evolving under the influence of noise. By considering the average
crossing time between two distinct trajectories, we show that
different Kauffman models exhibit a similar kind of behavior, even
when the structure of their basins of attraction is quite
different. This can be considered as a robust property of these
models.  We present numerical results for the full range of noise
level and obtain approximate analytic expressions for the above
crossing time as a function of the noise in the limit cases of small
and large noise levels.
\end{abstract}


\noindent {\bf Running title}: Kauffman Model with Noise.

\pagebreak

%%%%%%%%%%%%%%%%%%%%%%%%%%%%%%%%%%%%%%%%%%%%%%%%%%%%%%%%%%%%%%%%%
\section{Introduction}

The Kauffman model (or $N$-$K$ model) describes the dynamics of a
network of $N$ Boolean spins, each controlled by $K$ other spins
through a binary function. It was first proposed by S. A. Kauffman in
1969~\cite{69.1} as a model for cell differentiation and
genetic networks. Since then, its application has been extended to
many other fields in physics, biology, computational and social
sciences. During the past few decades, most of the work done on
Kauffman models has been dedicated to the study of the configuration
space structure, the length and number of cycles, the size of basins
of attraction, and the phase transition between ordered and disordered
phases (for references see~\cite{02.3}).  These properties are
obtained by considering the deterministic dynamics of the system,
which is well known by now. However, those studies have shown that
some of the generic properties of the Kauffman model are far from
being robust. The non-robustness of the deterministic dynamics is
reflected, for example, in the fact that by slightly changing a given
initial configuration of spins, the system may ``jump'' from one basin
of attraction to a very different one. On the other hand, due to the
exponential growth of the state space with $N$, it is often necessary
to thoroughly probe the state space in order to determine a generic
property of the system, such as the mean number of different basins of
attraction or the mean cycle length.  Actually, it has recently been
shown that a systematic bias due to an under-sampling of the state
space can be present in some of the results reported in the literature
during the last 30 years~\cite{01.7}.  Therefore it is
valuable to find a method which reveals the robust properties of the
Kauffman model.

Real networks are always subjected to external fluctuations.
Consequently, the relevant properties characterizing the network
should exhibit a certain degree of robustness to external
perturbations. In 1989, both Miranda \emph{et al}.~\cite{89.3} and
Golinelli \emph{et al}.~\cite{89.5} analyzed the stochastic dynamics
of the Kauffman model in the case in which an external noisy signal is
present.  In this work we extend the study of the stochastic dynamics
of the Kauffman network with noise, by performing more accurate
numerical simulations as well as analytic calculations. We focus our
attention on the time it takes for two trajectories, starting out from
different initial conditions, to cross. We consider two cases. First
the situation in which each one of the $N$ spins is determined by $K$
other spins chosen randomly from everywhere in the system (the
Kauffman net). The second case is a d-dimensional lattice in which
each spin is preferentially coupled to its immediate neighbors. As we
will see, both models exhibit qualitatively similar behavior.

In section \ref{sec:model}, we introduce the Kauffman model with
deterministic and stochastic dynamics. In section
\ref{sec:numerical} we describe our numerical results for both
Kauffman nets and lattices. In section \ref{sec:theoretical}, we
study closely the behavior of these models in the limits of small
and large noise. We summarize this work in section
\ref{sec:conclusions} with a brief discussion of the results.

%%%%%%%%%%%%%%%%%%%%%%%%%%%%%%%%%%%%%%%%%%%%%%%%%%%%%%%%%%%%%%%%%%%
\section{The Kauffman model}\label{sec:model}

%=============================================================
\subsection{Deterministic dynamics}
%=============================================================

A Kauffman model consists of $N$ Boolean spins $\{S_1,S_2,\dots,S_N\}$
with $S_i$ being either zero or one. The value of each spin $S_i$ at
time $t+1$ is determined by the values of $K$ other spins
$S_{i_1},S_{i_2},\dots, S_{i_K}$, which are called the {\em
controlling elements} for spin $S_i$. (The number $K$ is called the
\emph{connectivity} of the system.) Once the connections in the system
are established, each spin $S_i$ is assigned with a Boolean function
$f_i$ of its $K$ controlling elements. A realization of the Kauffman
model consists of the set of connections and Boolean functions
assigned to every spin. The dynamics of the network is then given by
\begin{equation}
S_i(t+1)=f_i(S_{i_1}(t),S_{i_2}(t),\ldots,S_{i_K}(t))\ \mbox{for
$i=1,\ldots,N$}. \label{eq:definition}
\end{equation}
For convenience, we will denote by $\Sigma_t$ the state of the
system at time $t$:
\[
\Sigma_t=\{S_1(t),S_2(t),\dots,S_N(t)\}.
\]
In different Kauffman models, the assignments of the $K$ controlling
spins $S_{i_1},S_{i_2},\dots, S_{i_K}$ of each spin $S_i$ and the
dynamic rules $f_i$, are different. In \emph{Kauffman nets}, the
controlling elements of $S_i$ are assigned randomly, whereas in a
\emph{Kauffman lattice} they are chosen only among its nearest neighbors. The
dynamic rules $f_i$ are chosen randomly in such a way that its two
possible outcomes, 0 and 1, occur with probability $\rho$ and $1-\rho$
respectively. If the realization of the network is time-independent,
the network is called \emph{quenched}, while if either the set of
connections or the set of Boolean functions $f_i$ are re-assigned at
every time step, the network is termed \emph{annealed}.

Annealed models are more convenient for theoretical studies than
quenched models. For example, by using the annealed approximation it
has been shown analytically that Kauffman nets exhibit three different
phases: frozen, critical and chaotic, depending upon the values of the
parameters $K$ and $\rho$ \cite{86.3}. The critical value of the
connectivity is given by $K_c=[2\rho(1-\rho)]^{-1}$. For $K<K_c$ the
system is in the frozen phase, whereas if $K>K_c$ it is in the chaotic
phase.  Throughout this work we will use $\rho=1/2$, for which
$K_c=2$.

But for most real cases (neural networks, genetic networks, etc.),
quenched models will be more appropriate since in real networks
neither the connections nor the interactions between the elements
change randomly at every moment. However, it has been shown that in
the limit $N\rightarrow\infty$, both the quenched and the annealed
Kauffman nets are exactly equivalent with respect to the evolution of
the overlap between different configurations, although not with
respect to the configurations themselves
\cite{86.1,87.3,87.6}.  In this paper, our main focus is on quenched
Kauffman nets and lattices.

Due to the finite size of the system, there are a finite number of
possible configurations, to wit $\Omega = 2^N$. Therefore,
starting out with an initial configuration, the system will
eventually fall into a previously visited state, after which the
same sequence of states repeatedly occurs again. The state space
breaks up into a multitude of cycles (or attractors). The totality
of points which end up in the same attractor represents its
\emph{basin of attraction}.

%===============================================================
\subsection{Stochastic dynamics}
%===============================================================

There are different ways of introducing noise into Kauffman models and
they reveal different features of the configuration space of the
model. Following Miranda and Parga \cite{89.3}, we introduced
noise in the following way:
\begin{equation}
S_i(t+1)=\left\{
\begin{array}{ll}
f_i(S_{i_1}(t),S_{i_2}(t),\ldots,S_{i_K}(t))&\mbox{with probability $1-r$},\\
1-f_i(S_{i_1}(t),S_{i_2}(t),\ldots,S_{i_K}(t))&\mbox{with
probability $r$}.
\end{array}
\right. \label{eq:noisedefinition}
\end{equation}
In this way, every spin $S_i$ has a probability $r$ of violating
the deterministic rule~(\ref{eq:definition}). We will say that an
\emph{$n$-spin flip event} has occurred at a particular time step,
if $n$ spins violated the deterministic rule in this time step.
Notice that this stochastic dynamic rule has a symmetry about the
point $r=0.5$. For $r>0.5$, if we make a substitution $f_i\rightarrow
1-f_i$, the rule becomes identical with the case of $1-r$. Since the
Boolean functions $f_i$ are assigned randomly, $f_i$ and $1-f_i$ are
equally likely to appear in a particular realization of the
model. After averaging over different realizations, the cases with
probabilities $r$ and $1-r$ are indeed identical. Due to this symmetry
in the stochastic dynamical rule~(\ref{eq:noisedefinition}), we only
need to consider the case $r\in[0,0.5]$. Note that for the particular
value $r=0.5$, $S_i(t+1)$ is equally likely to be zero or one
independently of the value of $f_i$.

In the presence of noise, the concept of ``attractor'' does not
hold any more, since as the system evolves, there is a non-zero
probability of ``jumping'' to a different attractor, and
consequently every point in the state space can be reached from
any initial condition. In this sense, the ``boundaries'' between
different attractors become more diffuse as the level of noise
increases \cite{87.5}. However, we shall argue that the system
has a sort of effective attractor even in the presence of noise,
specifically when $r$ is  large.

One of the interesting things to study is the time it takes for
two trajectories to cross. Suppose that we start with two
different initial configurations, $\Sigma_0$ and
$\tilde{\Sigma}_0$, and let them evolve according to
(\ref{eq:noisedefinition}), noting all the configurations
produced:
\begin{eqnarray}
\Sigma_0&\rightarrow&\Sigma_1\rightarrow\Sigma_2\rightarrow\dots
\rightarrow\Sigma_\tau\nonumber\\
\tilde{\Sigma}_0&\rightarrow&\tilde{\Sigma}_1\rightarrow\tilde{\Sigma}_2
\rightarrow\dots\rightarrow\tilde{\Sigma}_\tau\nonumber
\end{eqnarray}
The crossing time $\tau$ is then defined as the time for which either
one of the trajectories coincides for the first time with a
configuration previously visited by the other trajectory. For example,
when $\tilde{\Sigma}_\tau$ is equal to any of the configurations
$\Sigma_0,\Sigma_1,\Sigma_2,\dots,\Sigma_\tau$. Two important cases
have to be distinguished, when $\Sigma_0$ and $\tilde{\Sigma}_0$
belong to the same basin of attraction, and when they belong to
different basins of attraction. For those cases we will denote the
crossing time by $\tau_s$ and $\tau_d$, respectively.

Miranda and Parga examined the behavior of the system by considering
only the attractors with largest and next-largest basins of
attraction. They then showed that for small values of $r$, the
behavior of $\tau_s$ and $\tau_d$ are very different. At $r=0$, two
trajectories from the same basin will cross in a time comparable with
the sum of two times: first, the transient time required to enter the
attractor, and second, the length of the attractor itself. Conversely,
two trajectories starting out from different basins will never
cross. On the other hand, they found that for sufficiently large
values of $r$, the crossing time became independent of the starting
point. It did not matter where the two trajectories start, the two
basins merge into a sort of effective attractor and the trajectory
bounces around within that subset of the system-states. For these
larger values of $r$, the observed effective basin size increased with
$r$.  From this, they drew the conclusion that the disappearance of
basins of attraction with the increase of $r$ is a sort of
hierarchical process in the sense that in a finite period of time, the
portion of the whole state space explored by a trajectory starting out
from a given basin of attraction increases with $r$. Complete
randomness is achieved at $r=0.5$, where the trajectory explores the
entire state space.

As we will see, our simulation will show the same general behavior as
described by Miranda and Parga. But we shall explore the behavior in
more detail, showing the crossing time for the whole range of values
of $r$ and $K$.

%%%%%%%%%%%%%%%%%%%%%%%%%%%%%%%%%%%%%%%%%%%%%%%%%%%%%%%%%%%%%%%%%%%
\section{Numerical results}\label{sec:numerical}

Kauffman nets and Kauffman lattices differ in the structure of their
basins of attraction.  One would expect this difference to be
reflected in the response of these models to the influence of
noise. For random realizations of the coupling functions $f_i$, what
determines the basin structure is the connectivity $K$. Therefore, we
will first analyze separately the cases with large $K$ (chaotic phase)
and small $K$ (ordered phase).

We will partially follow Miranda and Parga's approach in that we
compute the average crossing time $\tau_s$ by using two initial
configurations, $\Sigma_0$ and $\tilde{\Sigma}_0$, in the largest
basin of attraction.  For the average crossing time $\tau_d$, we pick
one starting configuration in the largest basin and the other in the
next largest one. The reasons to choose only the two largest basins
of attraction will be clear in what follows.

%=======================================================
\subsection{Kauffman models with large $K$}
%=======================================================

\begin{figure}[h]
\psfig{file=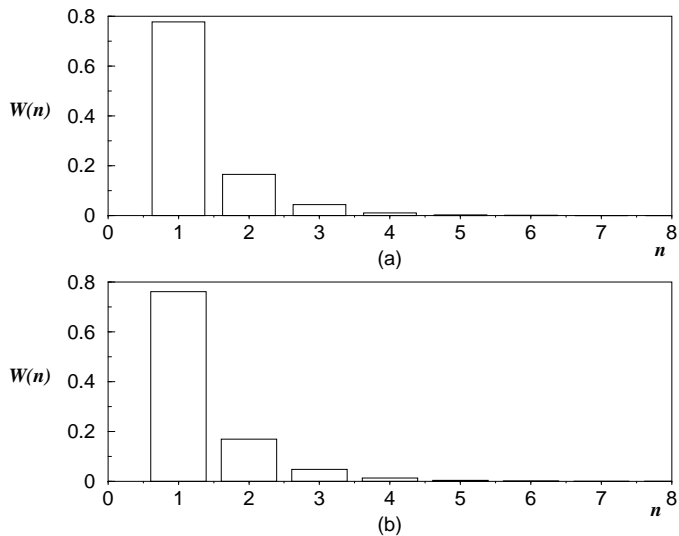,width=3.5in,clip=} \caption{
\small Distribution $W(n)$ of basin-sizes 
for: (a) the one-dimensional Kauffman lattice and (b) the Kauffman
net, both with $N=20$ and $K=5$. For the Kauffman net, the connections
between spins are chosen randomly, whereas for the lattice every spin
is connected to itself and to its $4$ nearest neighbors (periodic
boundary conditions were used). The number $n$ in the horizontal axis
corresponds to the $n$-th largest basin in the model.}
\label{fig:basin_size_largeK}
\end{figure}

We want first to characterize the structure of the basins of
attraction. One way of doing it is by computing the distribution of
basin-sizes $W(n)$, which is the fraction of the state space $\Omega$
occupied by the $n$-th largest basin of attraction. In
Fig.~\ref{fig:basin_size_largeK} we show $W(n)$ for a Kauffman net and
a 1-dimensional Kauffman lattice, both with $N=20$ and $K=5$. It can
be seen from this figure that both models exhibit a very similar
structure in their basins of attraction in the sense that the basin
sizes are similar.  It is worth mentioning that in other aspects, like
orbit length or transient time\footnote{The transient time is the time
it takes before a trajectory enters the stable cycle.}, the basins of
attraction can still be very different in both models.

From Fig.~\ref{fig:basin_size_largeK} it also can be seen that the
largest and next largest basins occupy more that $90\%$ of the whole
state space. Therefore, to a good approximation it can be assumed that
the dynamics takes place mainly in these two largest basins.

%
%
%-----------------------FIGURE V ----------------------------------
%-----------------------FIGURE ^ -----------------------------------
%
%

%
%
%-----------------------FIGURE V ----------------------------------
\begin{figure}[h]
\psfig{file=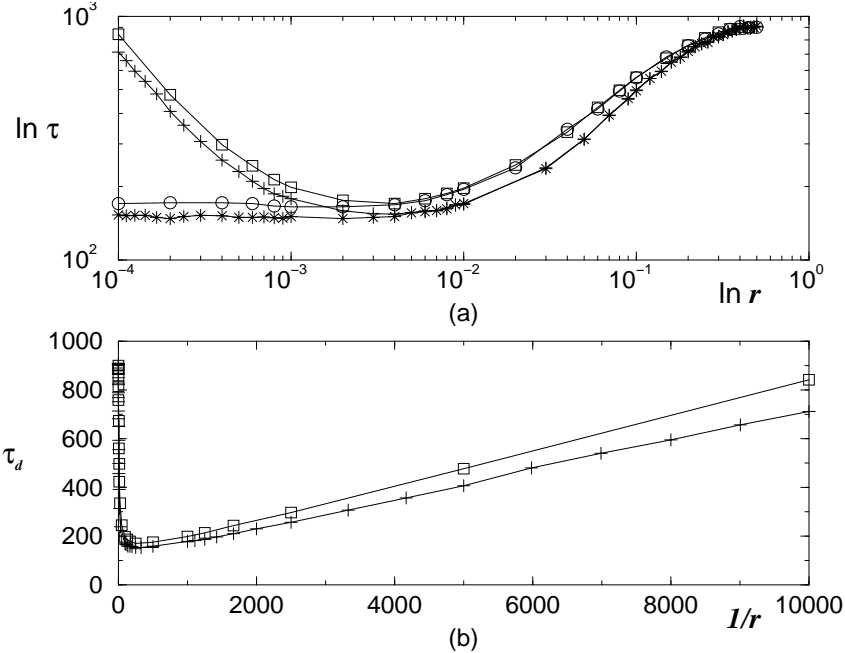,width=4.5in,clip=}
\caption{\small Average crossing time $\tau$ for different
Kauffman models with $N=20$ and connectivity $K=5$ (chaotic phase)
as a function of (a)~the noise intensity $r$ and (b)~the inverse of
the noise intensity. The symbols are as follows. $1$-dimensional Kauffman
lattice: ($\Box$) $\tau_d$ and ($\circ$) $\tau_s$. Kauffman net: ($+$)
$\tau_d$ and ($\ast$) $\tau_s$. Each point is the average over 4000
realizations of the model.}
\label{fig:averagetime_largeK}
\end{figure}
%-----------------------FIGURE ^ ----------------------------------
%
%

Fig.~\ref{fig:averagetime_largeK} shows the average crossing times
$\tau_s$ and $\tau_d$ as functions of $r$ for the Kauffman net and the
$1$-dimensional lattice both with $N=20$ and $K=5$ (chaotic
phase). Notice that these two kinds of $N$-$K$ models exhibit very
similar behavior under the influence of noise.

%=======================================================
\subsection{Kauffman models with small $K$}
%=======================================================

\begin{figure}[h]
\psfig{file=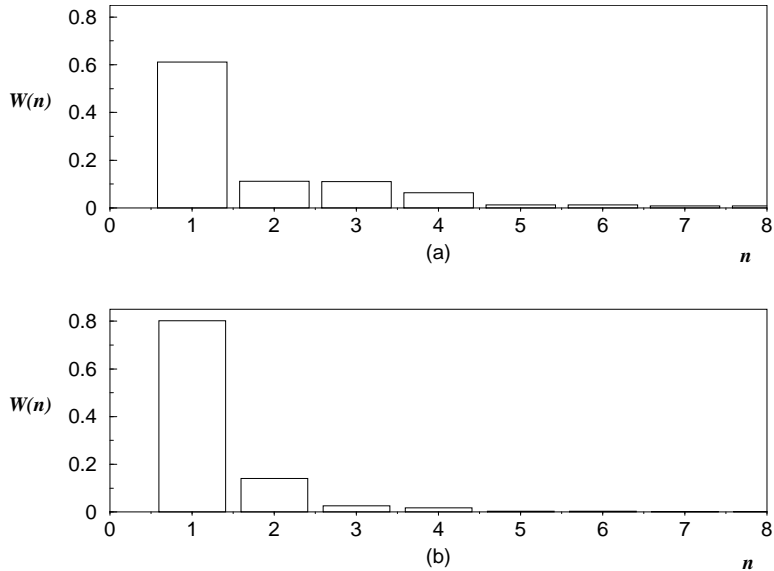,width=4in,clip=} 
\caption{
\small 
Distribution $W(n)$ of basin-sizes for: (a) the one-dimensional Kauffman
lattice and (b) the Kauffman net, both with $N=20$ and $K=1$(frozen
phase). The number $n$ in the horizontal axis has the same meaning as
in Fig.~\ref{fig:basin_size_largeK}.}
\label{fig:basin_size_smallK}
\end{figure}

The ordered phase is characterized by $K=1$ and $K=2$. In this
section we will present the results for the minimum value of $K$,
namely, $K=1$. In Fig.~\ref{fig:basin_size_smallK} we show $W(n)$
for a Kauffman net and a 1-dimensional Kauffman lattice, both
with $N=20$ and $K=1$. The connections in the Kauffman net were, as
usual, chosen randomly, whereas in the 1-dimensional lattice the
node $S_i$ was connected either to $S_{i-1}$ or to $S_{i+1}$ with
equal probability (we use periodic boundary conditions).

\begin{figure}[h]
\psfig{file=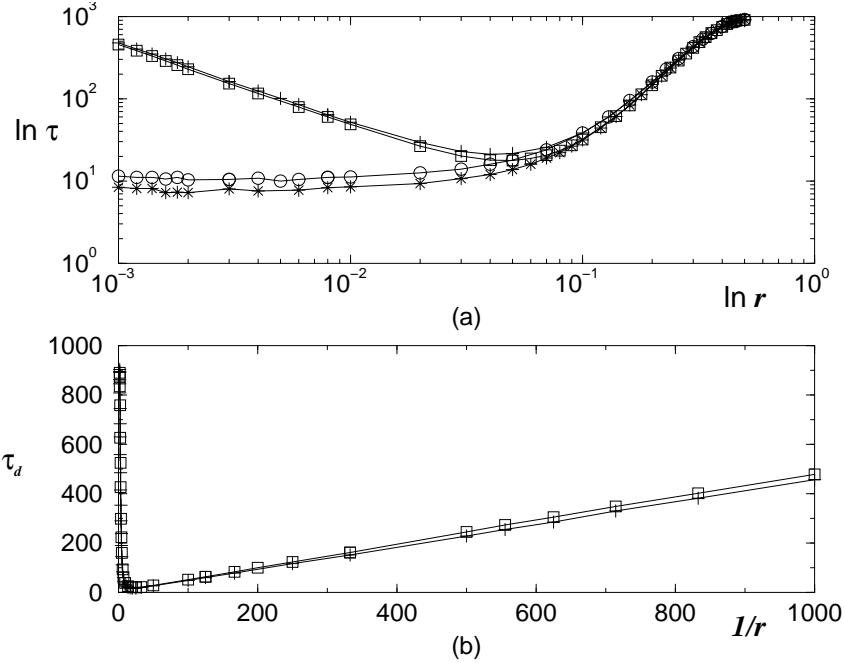,width=4.5in,clip=}
\caption{\small Average crossing time $\tau$ for different
Kauffman models with $N=20$ and connectivity $K=1$ (frozen phase) as a
function of (a)~the noise intensity $r$ and (b)~the inverse of the noise
intensity. The symbols are as follows. $1$-dimensional Kauffman
lattice: ($\Box$) $\tau_d$ and ($\circ$) $\tau_s$. Kauffman net: ($+$)
$\tau_d$ and ($\ast$) $\tau_s$. Each point is the average over 4000
realizations of the model.}
\label{fig:averagetime_smallK}
\end{figure}

From Fig.~\ref{fig:basin_size_smallK} it is apparent that in this
case, the basin structures of the Kauffman net and the Kauffman lattice are
less similar than in the chaotic phase. For the lattice, the two largest
basins of attraction no longer occupy more than 90\% of the whole state
space, whereas in the Kauffman net they still do. However, the response
to the influence of noise is mostly the same in both models, as can be seen
from Fig.~\ref{fig:averagetime_smallK} where the crossing times $\tau_s$
and $\tau_d$ are plotted as functions of the noise intensity $r$.

%
%
%-----------------------FIGURE V ----------------------------------
%-----------------------FIGURE ^ -----------------------------------
%
%

%
%
%-----------------------FIGURE V ----------------------------------
%-----------------------FIGURE ^ ----------------------------------
%
%

%================================================
\subsection{Robust behavior of the crossing time}
%================================================

Figures \ref{fig:averagetime_largeK} and \ref{fig:averagetime_smallK}
show that the behavior of the different Kauffman models under the
influence of noise have the following general characteristics, both in
the frozen and in the chaotic phases:

\begin{itemize}
\item For small $r$, $\tau_d$ decreases as $1/r$ while $\tau_s$ is
nearly constant.
\item For large $r$, both $\tau_d$ and $\tau_s$ increase with $r$ and 
become equal at $r=0.5$.
\item For intermediate values of $r$, $\tau_d$ has a minimum when
$\tau_s\approx\tau_d$.
\end{itemize}

Let us analyze separately each one of the above characteristics.

\begin{figure}[h]
\centering \psfig{file=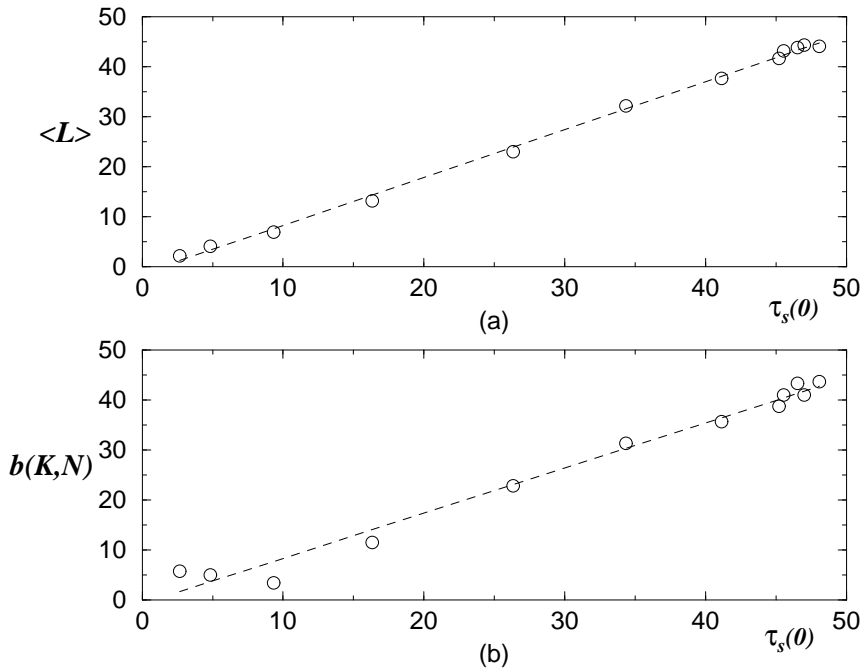,width=4.5in,clip=}
\caption{\small (a) Plot of the mean cycle length $\langle L \rangle$
as a function of $\tau_s(0)$ for a Kauffman net with $N=12$. Each
point corresponds to a different value of $K$, starting with $K=1$ for
the first point in the lower left corner of the graph and ending with
$K=12$ for the last point in the upper right corner. The dashed line
is the best linear fit to the numerical data (circles). (b) Same type
of graph as before but now showing the dependence of $b(K,N)$ on
$\tau_s(0)$. The slopes of the dashed lines are (a) $\sim 0.96$ and
(b) $\sim 0.90$, which shows that
$b(K,N)\approx\tau_s(0)\approx\langle L\rangle$. }
\label{fig:intercept}
\end{figure}

In the limit $r\rightarrow0$, $\tau_s$ approaches a finite value
$\tau_s(0)$, the mean crossing time for two trajectories in the
largest basin of attraction in the absence of noise. This crossing
time is roughly one half the average cycle length, plus one half the
average transient time.  The transient time and the cycle length are
of the same order of magnitude, therefore $\tau_s(0)$ is expected to
be approximately equal to the mean cycle length $\langle L \rangle$ of
the largest basin of attraction (see Fig.~\ref{fig:intercept}a).

On the other hand, $\tau_d$ diverges as $r\rightarrow0$. The
numerical data (see figures~\ref{fig:averagetime_largeK}b and
\ref{fig:averagetime_smallK}b) suggest that in this limit, $\tau_d$ has 
the form

\begin{equation}
\tau_d\approx {a(K,N)\over r}+b(K,N), \ \ \ \ r\rightarrow0.
\label{eq:1/r}
\end{equation}
The above divergence is due to the fact that, in the absence of
noise, there is a zero probability for a trajectory to jump
between different attractors. Under the deterministic dynamics,
every trajectory will remain within its own basin of attraction
for ever. In the next section we will see that the $1/r$ behavior
of $\tau_d$ is a consequence of the fact that the dynamics is
governed by one-spin flip events when $r$ is small.  For large
values of $K$, the largest basin occupies almost the whole state
space. Under these circumstances, every time a one-spin flip
occurs, the trajectory in the next largest basin will have a
finite probability of diverging very substantially from the path
it would have followed in the absence of noise. That divergence
will usually force the trajectory into the largest basin. In fact,
for a fraction of order one of the noise events in the smaller
basin, the noise will flip the trajectory into the largest one.
Once the two trajectories are in the largest basin, they have a
lifetime $b(K,N)$ before they cross. This lifetime is expected to
be of order one of $\tau_s(0)$, the typical length of the largest
basin of attraction. The above can actually be seen in
Fig.~\ref{fig:intercept}, from which it is apparent that for large
$K$, $b(K,N) \approx \tau_s(0) \approx \langle L \rangle$.
 
%
%
%-----------------------FIGURE V ----------------------------------
%-----------------------FIGURE ^ -----------------------------------
%
%

In the opposite limit $r\rightarrow0.5$, the crossing times $\tau_s$
and $\tau_d$ become equal, which means that for high levels of noise,
the barriers between different attractors become small. When $r$
reaches its maximum value $0.5$, all barriers vanish. In this case,
both trajectories randomly jump from one state to another throughout
the state space, and both $\tau_s$ and $\tau_d$ become equal to the
time it takes for two random walks to cross. As derived in
section~\ref{subsec:large_r}, this crossing time is the solution to
the ``birthday problem'', i.e.

\begin{equation}
\tau_s = \tau_d\propto 2^{N/2}
\label{eq:maximum_r}
\end{equation}

In this way we have obtained a qualitative description of the
limiting cases of figures~\ref{fig:averagetime_largeK} and
\ref{fig:averagetime_smallK}.  The one
qualitative feature left to describe is the crossover from the small
$r$ to the large $r$ behavior. As one can see from these figures, the
crossover occurs when the two times $\tau_s$ and $\tau_d$ become
roughly equal.  This in turn happens when
\begin{equation}
r \sim a(K,N)/ \tau_s(0) \sim 1/N
\label{eq:interim_r}
\end{equation}
Thus, the minimum in $\tau_d$ occurring between the two previous limit
values of $r$, can be interpreted as the result of a sort of
``competition'' between the randomness in the system (coming from the
presence of noise), and the barriers separating the attractors (which
come from the deterministic dynamics of the system).

%===================================================
\subsection{A Kauffman model with equal basin-sizes}
%===================================================

\begin{figure}[h]
\psfig{file=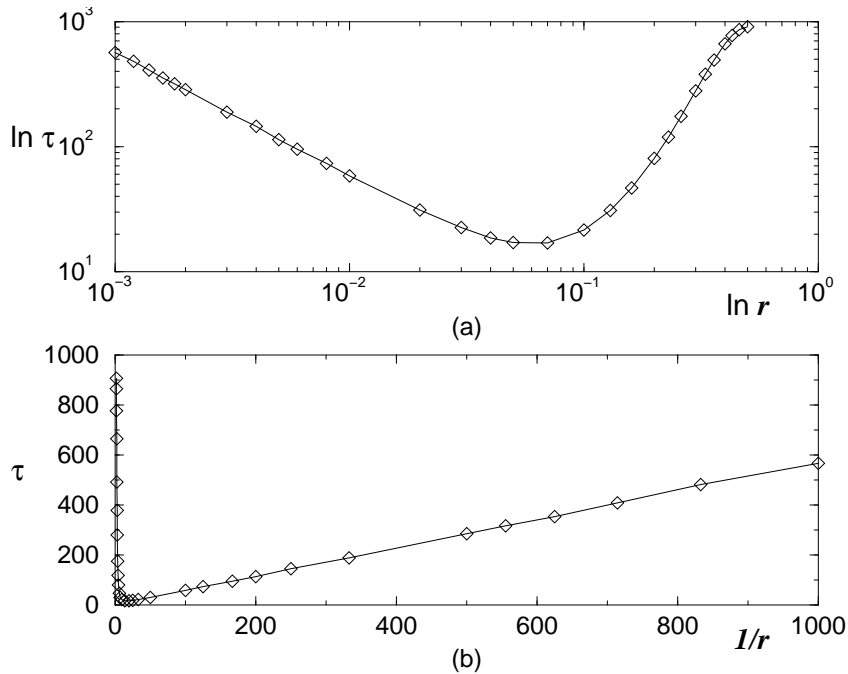,width=4.5in,clip=}
\caption{\small Average crossing time $\tau$($\Diamond$) for a
Kauffman lattice with $N=20$ and connectivity $K=1$ as a function
of (a)~the noise intensity $r$ and (b)~the inverse of the noise
intensity. For this model, every spin is correlated to itself and
we choose the Boolean functions in such a way that the whole state
space is composed of $8$ basins of attraction with equal size.
Each point is the average over $10000$ realizations of the model.}
\label{fig:averagetime_random}
\end{figure}

Finally, we would like to mention that the above results are also true
for Kauffman models in which all the basins of attraction have the
same weight. As an example, consider a Kauffman lattice with $N=20$
and $K=1$ in which every spin is connected to itself. For $K=1$ there
are only four Boolean functions $f_i(S)$: tautology $f_i(S)=1$,
contradiction $f_i(S)=0$, identity $f_i(S)=S$ and negation
$f_i(S)=1-S$. Imagine then the very specific realization in which two
of the coupling functions are identity, two are negation and all the
others are either tautology or contradiction.  By simple analysis, we
know that for this specific model, the whole state space is composed
of eight basins of attraction with equal
size. Fig.~\ref{fig:averagetime_random} shows the crossing time $\tau$
as a function of $r$ for this particular model. Since in this case all
the basins of attraction have the same size, the two initial
conditions needed to compute $\tau$ were chosen randomly among the
whole state space. Again, the $\tau\sim1/r$ behavior for small $r$ and
the $\tau\propto 2^{N/2}$ behavior for $r\rightarrow0.5$ are obtained.
Of course we can construct many other models by choosing different
coupling functions $f_i$ for this $K=1$ self-correlated case. All the
Kauffman models we have explored have shown this kind of behavior
under noise.

%
%
%-----------------------FIGURE V ----------------------------------
%-----------------------FIGURE ^ ----------------------------------
%
%

%%%%%%%%%%%%%%%%%%%%%%%%%%%%%%%%%%%%%%%%%%%%%%%%%%%%%%%%%%%%%%%%%%%%%%%%
\section{Theoretical analysis}\label{sec:theoretical}

%=======================================================
\subsection{Small $r$ limit} \label{subsec:small_r}
%=======================================================

In this region, the main characteristic of average crossing time is
that $\tau_d\sim 1/r$. The reason for this dependence is that the
stochastic dynamics is dominated by one-spin flip events, in the
following sense. The probability of a one-spin flip event ($\sim r$)
is much larger than the probability of a two-spin flip event ($\sim
r^2$). If the probability to jump to a different basin of attraction
in a one-spin flip event is significantly different from zero, then
the dynamics will be dominated only by this kind of events. Even if it
was necessary to flip two spins to jump from one basin to another,
this process can be decomposed into two one-spin flip events occurring
sequentially, instead of being carried out at once in one two-spin
flip event.

As we have shown in section~\ref{sec:numerical}, the $1/r$ behavior is
present in a wide variety of Kauffman models. This in turn, implies
that the one-spin flip events dominate the dynamics for small values
of the noise. In this subsection, our goal is to derive an expression
for the coefficient $a(K,N)$ for Kauffman nets. To do so, we will make
the assumption that the dynamics takes place only in the two largest
basins of attraction. Although the $\tau_d\sim 1/r$ behavior is
generally true, as we have found, the preceding assumption is not true
for all Kauffman models, especially for those with small values of
$K$, but as we show below, it becomes more valid as $K$ increases. 

\begin{figure}[h]
\centering \psfig{file=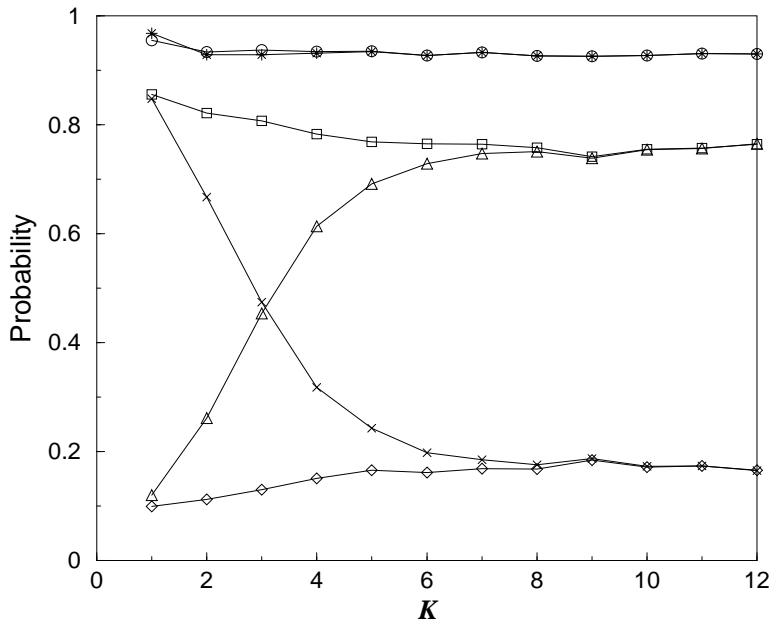,width=4 in,clip=}
\caption{\small Plot of the probabilities $P_{1,2}$ ($\Diamond$),
$P_{2,1}$ ($\triangle$), $Q_1$ ($\Box$) and $Q_2$ ($\times$) as
functions of $K$, for a Kauffman net with $N=12$. Also shown is
$M$ as a function of $K$, obtained as $M=P_{1,2}+Q_1$ ($\ast$),
and as $M=P_{2,1}+Q_2$ ($\circ$). Note that even though $M$ is a
constant for all values of $K$, the probabilities $P_{1,2}$ and
$P_{2,1}$ are rather small for small values of $K$. }
\label{fig:probabilities}
\end{figure}

To start the calculation of $a(K,N)$, let $P_{1,2}$ be the probability
for jumping from the largest basin to the next largest basin with a
one-spin flip event and $P_{2,1}$ be a similar probability but jumping
in the opposite direction\footnote{By definition, the ratio of
$P_{1,2}$ and $P_{2,1}$ is strictly the inverse of the ratio of the
size of the largest basin to the next largest one in one
realization.}.  Let us also define $Q_1$ and $Q_2$ as the
probabilities of remaining in the largest basin and in the next
largest one, respectively, after one-spin flip event. Simulations show
that for Kauffman nets these probabilities have a slight dependence on
$N$ but a very strong dependence on $K$. The result is that
$P_{1,2}+Q_1\approx P_{2,1}+Q_{2}\approx M$, where $M$ is
approximately constant for all values of $K$ and $M>0.92$ (see
Fig.~\ref{fig:probabilities}).

However, the dynamics depends not only on the total sum $M$ but
also on the particular values of $P_{1,2}$, $Q_1$, $P_{2,1}$ and
$Q_2$. For small values of $K$, both $P_{1,2}$ and $P_{2,1}$ are
very small and comparable with $(1-M)$. So, even though $M$ is
large, the interaction between the two largest basins is weak in
the sense that the probability of jumping into smaller basins is
of the same order as $P_{1,2}$ and $P_{2,1}$.  Therefore, for
small values of $K$ the smaller basins play a significant role in
the dynamics of the system. The above can be seen in
Fig.~\ref{fig:probabilities}, in which the probabilities
$P_{1,2}$, $P_{2,1}$, $Q_1$, $Q_2$, and the sum $M$ are plotted as
functions of $K$. From this figure it is apparent that the two
largest basins of attraction are the dominant ones for large
values of $K$ (say $K \geq 5$).  The closeness of $M$ to $1$ means
that a trajectory will seldom jump into a basin other than the two
largest ones with only one-spin flip event.  In view of this
result, in some of the arguments below we will assume that $K$ is
sufficiently large so that the dynamics takes place only in the
two largest basins of attraction.

%
%
%-----------------------FIGURE V ----------------------------------
%-----------------------FIGURE ^ -----------------------------------
%
%

With the information about these probabilities, we can give an
approximate calculation of $\tau_d$ as a function of $r$ for the
case in which $K$ is large. We know that every spin violates the
deterministic rule (\ref{eq:definition}) with probability $r$.
Therefore, the probability of a one-spin flip event is $Nr$ and
consequently the expected time for this event to occur is
$T={1/(Nr)}$.  This is true for all configurations. For
sufficiently small values of $r$, this expectation time is much
longer than the average crossing time for two configurations in
the same basin. Hence, once two configurations jump into the same
basin, their trajectories meet before the next spin-flip event
becomes possible.

\begin{figure}[h]
\psfig{file=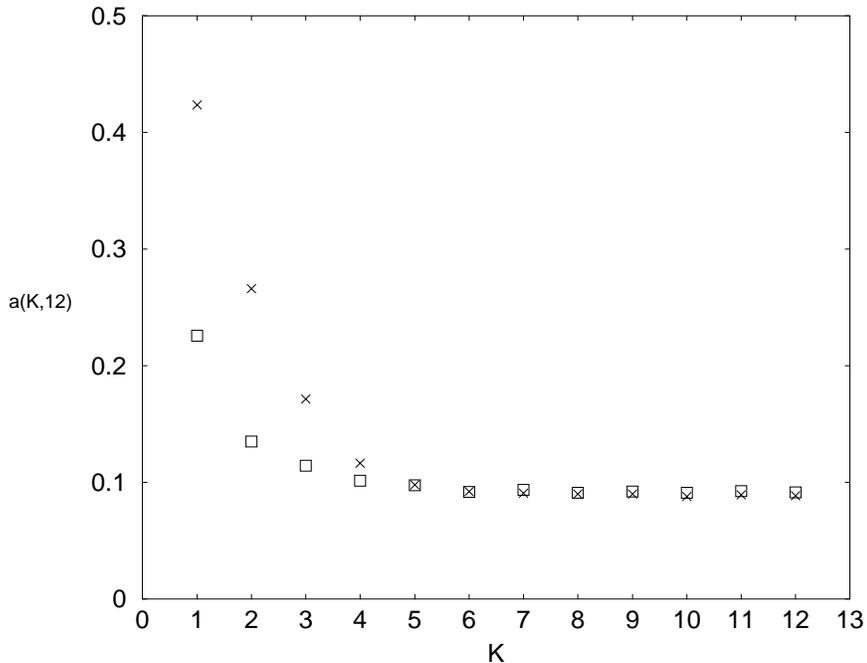,width=4.5in,clip=} \caption{The
coefficient $a(K,N)$ in Eqn.~(\ref{eq:1/r}) for different values of
$K$ for a Kauffman net with $N=12$. ($\times$) is the result obtained
from simulations by sampling the whole state space. ($\Box$) is the
result obtained from the theoretical result
Eqn.~(\ref{eq:coefficient1}).} \label{fig:coefficient}
\end{figure}

There are two cases in which the two trajectories meet after the
occurrence of a one-spin flip event at time $T$: the configuration in
the largest basin remains in it while the configuration in the next
largest basin jumps into the largest one, or vice versa. The above
occurs with probability $(Q_{1}P_{2,1}+Q_{2}P_{1,2})$. Similarly we
get the probability for the crossing of the two trajectories after
one-spin flip events at $2T$, $3T$, etc. The average value of $\tau_d$
is then:
\begin{eqnarray}
\tau_d & \approx & T\cdot (Q_{1}P_{2,1}+Q_{2}P_{1,2})+2T\cdot
(Q_{1}Q_{2}+P_{1,2}P_{2,1})\cdot (Q_{1}P_{2,1}+Q_{2}P_{1,2})+\cdots
\nonumber \\
&=&\sum_{m=1}^{\infty} m\cdot T\cdot
(Q_{1}P_{2,1}+Q_{2}P_{1,2})\cdot
(Q_{1}Q_{2}+P_{1,2}P_{2,1})^{m-1}\nonumber \\
&=&{1\over Nr}\cdot {Q_{1}P_{2,1}+Q_{2}P_{1,2}\over
(1-Q_{1}Q_{2}-P_{1,2}P_{2,1})^2}
\nonumber \\
&=& \frac{a(K,N)}{r}
\label{eq:coefficient}
\end{eqnarray}
where $a(K,N)$ is explicitly given by
\begin{equation}
a(K,N)=\frac{1}{N}\frac{Q_{1}P_{2,1}+Q_{2}P_{1,2}}
{(1-Q_{1}Q_{2}-P_{1,2}P_{2,1})^2}
\label{eq:coefficient1}
\end{equation}
In the above derivation, which is true for the case in which $K$
is large, we have ignored the time $b(K,N)$ it takes for two
trajectories to cross after they have jumped into the same basin.
Fig.~\ref{fig:coefficient} shows the coefficient $a(K,N)$ obtained
from equation~(\ref{eq:coefficient1}) and from simulation, for
different values of $K$ in a Kauffman net with $N=12$. It can be
seen that the simulation and the theoretical result agree very
well for $K\geq 5$. It is worth emphasizing that for small values
of $K$ and other Kauffman models where the two largest basins
don't have such dominance, the effect of other basins besides the
largest and next largest ones has to be considered to perform an
accurate derivation for the coefficient $a(K,N)$.

%
%
%-----------------------FIGURE V ----------------------------------
%-----------------------FIGURE ^ -----------------------------------
%
%

%==========================================================
\subsection{Large $r$ limit} \label{subsec:large_r}
%==========================================================

When $r$ acquires its maximum value $0.5$, the barriers between
different attractors vanish and the dynamics transforms into
a random mapping of the state space into itself
\cite{60.1,87.9}. In this limit, the crossing times $\tau_s$
and $\tau_d$ become indistinguishable and we will refer to both of
them simply as $\tau$. To understand the limit $r\rightarrow0.5$ we
will first give a simple ``birthday-problem'' argument to obtain the
order of magnitude of $\tau$. Then we will proceed to a more elaborate
analysis to obtain an approximate expression for $\tau$ as a function
of $r$, valid for $r$ close to $0.5$.

Imagine two walkers moving at random through a space of dimension
$\Omega=2^N$. They go through a number of steps $n=1, 2, 3\dots$ As
each step is completed, they have a total number of chances $C_n$ on
landing on some place previously covered by the other walker.  At the
first step we could have the two walkers at identical positions. Thus,
$C_1=1$.  At the second step, $C_2=3+C_1$ since each can land on the
original position of the other or they can both land at precisely the
same place.  After $n$ steps $C_n=C_{n-1} +2n+1= n^2$.  For large
$\Omega$ and small $n$, the probability of having collided with the
path is then of order $p_n \approx C_n/ \Omega$.  The average time for
crossing is then roughly given by the $n$-value for which $p_n$
becomes of order unity, so that $n^2$ is of order $\Omega$ or $n =
O(\Omega^{1/2})$. Therefore, the crossing time $\tau$ satisfies
$\tau\propto\Omega^{1/2}$.

To derive a more precise functional relation between $\tau$ and $r$ we
have to compute the probability $p_c(t)$ for two trajectories
$\{\Sigma_0,\Sigma_1,\dots, \Sigma_\tau\}$ and
$\{\tilde{\Sigma}_0,\tilde{\Sigma}_1,\dots, \tilde{\Sigma}_\tau\}$ to
cross at time $\tau$. In order to do that, we have to have first the
probability $p$ for two configurations $\Sigma_t$ and
$\tilde{\Sigma}_{\tilde{t}}$ to be the same (note that $t$ and
$\tilde{t}$ might be different). Let $S_i(t-1)$ be in $\Sigma_{t-1}$
and $\tilde{S}_i(\tilde{t}-1)$ be the corresponding spin in
$\tilde{\Sigma}_{\tilde{t}-1}$. Since these spins are in the same
position (each in its respective configuration), the deterministic
rule $f_i$ they obey is the same.  Notice that the dynamical equation
(\ref{eq:noisedefinition}) can be written as
\begin{equation}
S_i(t+1)=\left\{
\begin{array}{ll}
f_i&\mbox{with probability $1-2r$},\\
\mbox{evolve randomly}&\mbox{with probability $2r$}.
\end{array}
\right. \label{eq:noisedefinition_1}
\end{equation}
From this expression it follows that the probability for $S_i(t-1)$
and $\tilde{S}_i(\tilde{t}-1)$ to evolve according to the
deterministic rule $f_i$ is $(1-2r)^2$. Let us denote by $p_1$ the
probability that $S_i(t)=\tilde{S}_i(\tilde{t})$ when both spins are
updated according to the deterministic rule $f_i$.  To calculate $p_1$
we follow the annealed approximation introduced by Derrida and Pomeau
\cite{86.3}, which leads us to the following two possibilities:

\begin{enumerate}
\item
The $K$ inputs of $S_i(t-1)$ and $\tilde{S}_i(\tilde{t}-1)$ are the
same, which occurs with probability $1/2^K$. When this happens,
$S_i(t)=\tilde{S}_i(\tilde{t})$ with probability $1$.
\item
At least one of the inputs is different, which occurs with probability
$(1-1/2^K)$. In this case, if the evolution rules $f_i$ are assigned
in a sufficiently random way, there is a probability of 1/2 that
$S_i(t)=\tilde{S}_i(\tilde{t})$.
\end{enumerate}
From the above it follows that
\[
p_1=(1-2r)^2\left[\frac{1}{2^K}+\left(1-\frac{1}{2^K}\right)\cdot
\frac{1}{2}\right]
\]
On the other hand, the probability $p_2$ that $S_i(t)=\tilde{S}_i(\tilde{t})$
when the evolution rule $f_i$ is violated in one or both of the
configurations, is simply given by
\[
p_2=[1-(1-2r)^2]\cdot {1\over 2}
\]
Combining the values of $p_1$ and $p_2$ given above, the probability
$p$ for both configurations $\Sigma_{t}$ and $\tilde{\Sigma}_{\tilde{t}}$
to be equal is
\begin{eqnarray}
p&=&\left \{ (1-2r)^2\left [{1\over 2^K}+(1-{1\over 2^K})\cdot
{1\over 2}\right ]+[1-(1-2r)^2]\cdot {1\over 2}\right \} ^N\nonumber \\
&=&{1\over 2^N}\cdot \left \{ 1+{1\over 2^K}\cdot (1-2r)^2\right
\}^N \label{eq:p}
\end{eqnarray}
If $q(t)$ is the probability that the two trajectories have not yet
crossed at time $t$, then $p_c(t)$, the probability for the two
trajectories to cross at time $t$, is given by
\begin{equation}
p_c(t)=q(t-1)-q(t) \label{eq:p_c}
\end{equation}
The two trajectories are still separated at time $t$ if none of
the configurations $\{\Sigma_0,\Sigma_1,\dots,\Sigma_t\}$ is equal
to any of the configurations
$\{\tilde{\Sigma}_0,\tilde{\Sigma}_1,\dots,\tilde{\Sigma}_{t}\}$.
The probability for this to happen is
\[
q(t)=(1-p)^{(t+1)^2}
\]
Substituting this value of $q(t)$ into equation (\ref{eq:p_c}) we
get
\begin{equation}
p_c(t)= (1-p)^{t^2}-(1-p)^{(t+1)^2}
\end{equation}
Therefore, the average crossing time $\tau=\sum_{t=1}^{\infty}t\cdot
p_c(t)$ is given by
\begin{eqnarray}
\tau&=&\sum_{t=1}^{+\infty}t\cdot[(1-p)^{t^2}-(1-p)^{(t+1)^2}]\\
&\approx& \int_0^{+\infty}-t{d(1-p)^{t^2}\over dt}dt \nonumber \\
&=&{1\over 2}\sqrt{\pi\over -\ln{(1-p)}} \nonumber
\label{theory1}
\end{eqnarray}
Expanding the logarithm in the above equation around $p=0$, and
retaining only the terms up to the the first order, we finally get
\begin{equation}
\tau \approx {\sqrt{\pi} \over 2}\left [ 2\over 1+{1\over
2^K}\cdot (1-2r)^2 \right ]^{N/2} \label{eq:tau}
\end{equation}

It can be seen that Eqn.~\ref{eq:tau} is consistent with the
``birthday-problem'' argument for the case $r=0.5$.  When $r$ is not
exactly $0.5$, we do not have a simple birthday problem because the
coupling between different elements, and consequently the functions
$f_i$, still play a role in the dynamics. However, it is clear from
the above equation that the problem can be viewed as a birthday
problem with an effective state space $\Omega_{eff}=\left [ 2\over
1+(1-2r)^2/2^K \right ]^{N}$.

%
%
%-----------------------FIGURE V ---------------------------------
\begin{figure}[h]
\psfig{file=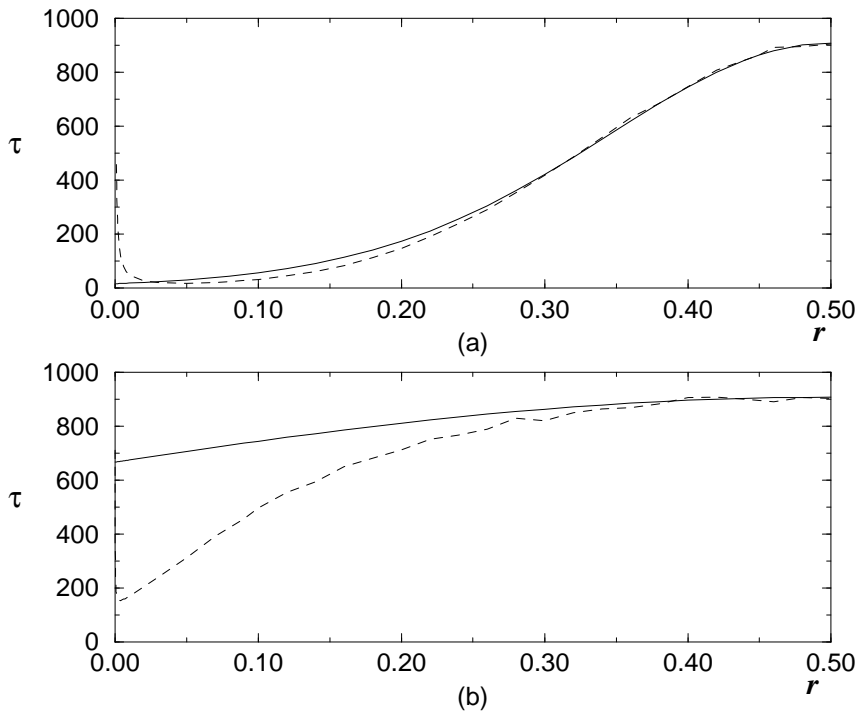,width=4.5in,clip=} \caption{ Average
crossing time $\tau$ as a function of $r$ for two Kauffman nets with
$N=20$ and connectivities (a) $K=1$ and (b) $K=5$. In both graphs the
dashed curve is the result of the numerical simulation for $\tau_d$,
whereas the solid line is the plot of the analytic expression
(\ref{theory1}).} \label{fig:theory_t}
\end{figure}
%-----------------------FIGURE ^ ----------------------------------
%
%

Fig.~\ref{fig:theory_t} compares the theoretical result
(\ref{eq:tau}) with the numerical simulation for $K=1$ and $K=5$.
As can be seen, the analytic result approximates very well the
numerical data in the region of $r$ close to $0.5$. In this
region the annealed approximation holds because the noise
breaks the correlations between the spins. However, for small
values of the noise those correlations are important and cannot
be neglected. We therefore do not expect agreement between the
numerical and theoretical results for small values of $r$ since
in this region the annealed appriximation is not longer valid.

%%%%%%%%%%%%%%%%%%%%%%%%%%%%%%%%%%%%%%%%%%%%%%%%%%%%%%%%%%%%%%%%%%%
\section{Conclusions}\label{sec:conclusions}

We have considered the effect of external perturbations (noise) in the
dynamics of the Kauffman model. The behavior of both, the Kauffman net
and the Kauffman lattice under the influence of noise is very similar,
even though these models might have a quite different structure in
their basins of attraction. In this sense, the response of the
Kauffman models to the effect of noise can be considered as a very
robust property.

In the limit $r\rightarrow0$, the most important property is the $1/r$
behavior of the crossing time $\tau_d$, which has been always present
in the Kauffman models we have studied so far. This $1/r$ behavior is
a consequence of the fact that, for small values of the noise, the
dynamics is dominated by one-spin flip events. An approximate equation
relating $\tau_d$ and $r$ was obtained by taking into account the fact
that, for large values of the connectivity, most of the dynamics takes
place in the largest and next largest basins of attraction.

In the second limit $r\rightarrow0.5$, the barriers between different
attractors disappear and the dynamics transforms into a random mapping
of the state space into itself. As a consequence, $\tau_s$ and
$\tau_d$ become equal. In the case in which $r=0.5$, the crossing time
between the two trajectories can be seen as the solution of a
``birthday problem'' in a space of size $\Omega=2^N$. For other values
of $r$, but still close to $0.5$, the correlations between spins have
to be taken into account, which have the effect of reducing the size
of the region of the state space explored by the dynamics.

Between these two limit cases for the noise, there is a minimum in the
value of $\tau_d$ as a function of $r$.  In a loose sense, this minimum
could be interpreted as the result of a ``competition'' between the
randomness generated by the noise, which tends to homogenize the state
space by diminishing the barriers across different attractors, and the
deterministic dynamics, which tends to confine two trajectories within
the same basin. To analyze this region it would be necessary to
consider multiple-spin flip events as well as long-time step
correlations.

The results and techniques presented in this work could be extended to
other systems acting under the influence of noise in order to provide
them with a robust characterization.

\subsection*{Acknowledgements}
This work was supported in part by the MRSEC Program of the National
Science Foundation under award number 9808595, and by the NSF DMR
0094569. We also thank to the Santa Fe Institute of Complex Systems
for partial support through the David and Lucile Packard Foundation
Program in the Study of Robustness.

%%%%%%%%%%%%%%%%%%%%%%%%%%%%%%%%%%%%%%%%%%%%%%%%%%%%%%%%%%%%%%%%%%%

\pagebreak

\bibliographystyle{plain}
\bibliography{bibliography}

\end{document}